\documentclass[aps,prl,twocolumn,groupedaddress,superscriptaddress,showkeys]{revtex4}
\usepackage{graphicx}
\usepackage{amsmath}

\begin{document}


\title{Dynamical effects of an unconventional current-phase relation in
YBCO dc-SQUIDs}

\author{T. Lindstr\"om}
\email[]{tobiasl@fy.chalmers.se}
\affiliation{Department of Microelectronics and Nanoscience,
Chalmers University of Technology and G\"oteborg University,
SE-412 96 G\"oteborg, Sweden}

\author{S.A. Charlebois}
\affiliation{Department of Microelectronics and Nanoscience,
Chalmers University of Technology and G\"oteborg University,
SE-412 96 G\"oteborg, Sweden}

\author{A.Ya. Tzalenchuk}
\affiliation{National Physical Laboratory, Teddington,Middlesex,
TW11 0LW, UK}

\author{Z. Ivanov}
\affiliation{Department of Microelectronics and Nanoscience,
Chalmers University of Technology and G\"oteborg University,
SE-412 96 G\"oteborg, Sweden}

\author{M.H.S. Amin}
\affiliation{D--Wave Systems Inc., 320-1985 Broadway, Vancouver,
B.C., V6J 4Y3, Canada}

\author{A.M. Zagoskin}
\affiliation{D--Wave Systems Inc., 320-1985 Broadway, Vancouver,
B.C., V6J 4Y3, Canada}
\affiliation{Physics and Astronomy Dept.,
The University of British Columbia, 6224 Agricultural Rd.,
Vancouver, V6T 1Z1 Canada}

\date{\today}

\begin{abstract}
The predominant $d$-wave pairing symmetry in high temperature
superconductors allows for a variety of  current-phase relations
in Josephson junctions, which is to a certain degree fabrication
controlled. In this letter we report on direct experimental
observations of the effects of a non-sinusoidal current-phase
dependence in YBCO dc-SQUIDs, which agree with the theoretical
description of the system.
\end{abstract}


\maketitle


It is well established \cite{tsuei} that the wave function of a
Cooper pair in most cuprate high-temperature superconductors (HTS)
has a $d$-wave symmetry. Its qualitative distinction from e.g. the
anisotropic s-wave case is that the order parameter changes sign
in certain directions, which can be interpreted as an {\em
intrinsic} difference in the superconducting phase between the
lobes equal to $\pi$.

The latter leads to a plethora of effects, like formation of
Andreev bound states at surfaces and interfaces in certain
crystallographic orientations \cite{Alff,Lofwander2001,Hu}. The
current-phase dependence $I_S(\phi)$ in Josephson junctions
formed by {\em dd}-junctions, as well as by {\em sd}-junctions
comprised of a cuprate and a conventional superconductor, depends
both on the spatial orientation of the d-wave order parameter
with respect to the interface, and on the quality of the latter
\cite{Yip,Zagoskin,ilichev99,ilichev2001,komissinskii}.
Time-reversal symmetry can also be spontaneously violated and
thus spontaneous currents generated
\cite{Huck,Omelyanchouk,ostlund}. Another effect can be doubling
of the Josephson frequency \cite{Zagoskin,Wendin,Arie}.

In this letter we report on experimental observations of strong
effects of an unconventional current-phase relation on the
dynamics of two $dd$-junctions integrated into a superconducting
interference device (SQUID) configuration.

Since $I_S(\phi)$ must be a $2\pi$-periodic odd function, it can
be expanded in a Fourier series. In the most cases only the first
two harmonics give a significant contribution to the current:
\begin{equation}
I_S(\phi)=I_c^I\sin \phi -I_c^{II}\sin 2\phi \label{eq:cpr}
\end{equation}
In Josephson systems of conventional superconductors the second
harmonic will usually be negligible \cite{Keene} but in {\em
dd}-junctions the second harmonic may dominate. If
$I_c^{II}>I_c^{I}/2$ the equilibrium state is no longer $\phi=0$
but becomes double degenerate at $\phi=\pm \arccos
(I_c^{I}/2I_c^{II}) \rightarrow \pi/2$. The system can then
spontaneously break time reversal symmetry by choosing either
state. Spontaneous currents as well as fluxes can be generated in
this state. The potential will have the shape of a double well and
there are reasons to believe that it will be possible to observe
quantum-coherence in this system. The presence of a
2$^{\text{nd}}$ harmonic in the current-phase relation(CPR) of a
{\em dd}-junction was confirmed by Il'ichev et al.
\cite{ilichev2001}.

A non-sinusoidal CPR of the junctions will change the dynamics of
a dc-SQUID \cite{ACR}. Regarding the junctions as magnetically
small, the supercurrent through the SQUID in the presence of an
external flux $\Phi_x\equiv \Phi_0 \cdot (\phi_x/2\pi)$ can be
written as
\begin{multline}
I_s(\phi,\phi_x)=I_{c1}^{I} \sin \phi -I_{c1}^{II} \sin(2\phi)+
I_{c2}^{I} \sin (\phi+\phi_x)
\\-I_{c2}^{II} \sin 2(\phi+\phi_x) \label{eq:Iph}
\end{multline}
 The critical current through the SQUID is given by the usual
expression $I_c(\phi_x)=\max_{\phi} I_s(\phi,\phi_x)$. The
time-averaged voltage over the SQUID in the resistive regime is
readily obtained in the resistively shunted junction (RSJ)
approximation. By introducing $\delta[\phi,\phi_x] =
\phi_2-\phi_1\label{eq:delta_definition}$ and applying the same
method as in \cite{Barone} with the necessary generalizations, we
obtain for the average voltage over the SQUID:
\begin{eqnarray}
\bar{V}^{-1} &=& {G_1+G_2 \over 2\pi} \int_{-\pi}^{\pi}d \phi
\left[ I- (G_1-G_2)\frac{\hbar}{2e}\frac{d\delta}
{dt}\right. \nonumber\\
&& \left.  -I_1 \left(\phi+{\delta \over 2} \right) - I_2
\left(\phi-{\delta \over 2} \right) \right]^{-1}. \label{eq:Vbar}
\end{eqnarray}

Here $G_{1,2}$ are the normal conductances of the junctions, and
\begin{eqnarray}
 \delta + \phi_x +
\frac{\pi L}{\Phi_0}(I_2(\phi-\delta/2)-I_1(\phi+\delta/2)) =
0\label{eq:delta_implicit},
\end{eqnarray}
gives the difference, $\delta$, in phase drops across each
junction. In deriving (\ref{eq:Vbar},\ref{eq:delta_implicit}) we
have assumed that the inductance $L$ is equally divided between
the SQUID arms. We have also neglected the spontaneous magnetic
fluxes in the $dd$-junctions, due to their small amplitude
\cite{Omelyanchouk,amin2}. Though (\ref{eq:delta_implicit}) is
only explicitly solvable in the limit $L\to 0$, it always yields
$\delta[-\phi,-\phi_x]=-\delta[\phi,\phi_x]$. This means that the
usual inversion symmetry is retained.

\begin{figure}
\includegraphics[width=7.5cm]{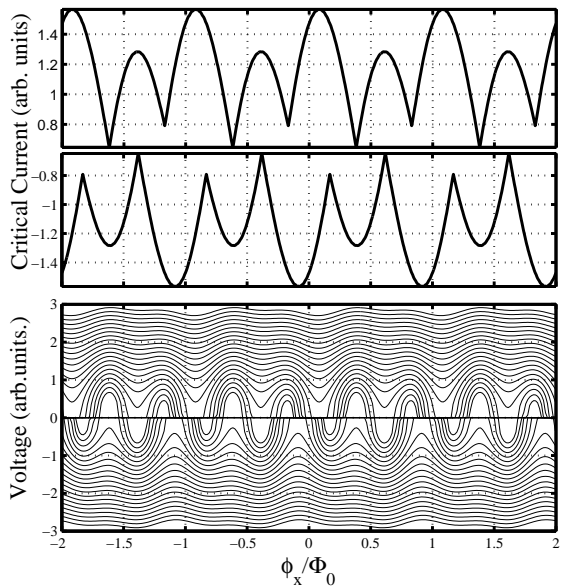}
\caption{The results of simulations
 of the $I_c-\phi_x$ and $\bar{V}-\phi_x$ dependence
for a dc-SQUID with $I_{c1}^I=1$, $I_{c2}^I=0.1$,
$I_{c1}^{II}=0.2$ and $I_{c2}^{II}=0.4$ (arb. units). The
different curves correspond to bias currents in the range
$I=I_{c1}^I$ to $I=5I_{c1}^I$.We assume $L=0$ and $G_1=G_2$. }
\label{fig:simulations}
\end{figure}

The results of numerical calculations based on (\ref{eq:Iph}) and
(\ref{eq:Vbar}) are shown in fig.\ref{fig:simulations}. The cusps
in the critical current correspond to the points at which the
global maximum in (\ref{eq:cpr}) switches from one local maximum
to another \cite{ACR}. Note the quasi-$\Phi_0/2$-periodicity of
the current isolines  in the $\bar{V}-\phi_x$ picture, reflecting
the current-phase dependence (\ref{eq:cpr}), and their shift along
the $\Phi_x$-axis, which depends on the sign of the bias current
(as it must to maintain the central symmetry with respect to the
origin). The shift does not depend on the magnitude of the current
since we neglect the self-inductance. For large biases the
$\Phi_0$-periodicity is restored. Indeed, as the bias grows, one
set of minima of the washboard potential, $U= (h/2e)[-I^I\cos\phi+
(I^{II}/2) \cos 2\phi - I\phi]$, disappears first unless the first
harmonic $I^{I}$ is {\em exactly} zero.


We have fabricated and studied a large number of dc-SQUIDs. The
samples were fabricated from 250 nm thick YBCO-films deposited on
SrTiO$_3$-bicrystals. The grain-boundary junctions (GBJs) are of
the asymmetric [001]-tilt type with the misorientation angle of
45$^\circ$ ($0^\circ-45^\circ$ GBJ). For more information on GBJs
see for example reference \cite{hilgenkamp}.

The pattern was defined using E-Beam lithography and then
transferred to a carbon mask employing a multi-step process.
Finally, the YBCO is etched through the mask using ion-milling.
This scheme allows us to fabricate high-quality bicrystal
junctions as narrow as 0.2 $\mu$m, as has been reported elsewhere
\cite{quox2002}. In the SQUIDs under investigation the junctions
are nominally 2 $\mu m$ wide; hence the fabrication-induced
damage of the junctions is small.

The measurements were done in an EMC-protected environment using a
magnetically shielded LHe-cryostat. However, the magnetic
shielding is imperfect, as is evident from the fact, that the
expected zero-field response of our SQUIDs is not exactly at zero.
The measuring electronics is carefully filtered and
battery-powered whenever possible. In order to measure the
dependence of the critical current on the applied field we used a
voltage discriminator combined with a sample-and-hold circuit. All
measurements reported here were performed at 4.2K.

The SQUID loops are 15$\times$15 $\mu$m$^2$. The numerically
calculated inductance \cite{khapaev} is approximately 25 pH,
yielding the factor $\beta=2\pi LI_c/\Phi_0$ between 0.5-2.

The SQUIDs were largerly non-hysteretic with a resistance of about
2 $\Omega$. The measured critical current varies from sample to
sample but is in the range of tens of microamperes giving a
current density of the order of $J_c=10^3$ A/cm$^2$. The estimated
Josephson penetration length $ \lambda_J=\Phi_0/\sqrt{
4\pi\mu_0j_c\lambda_L}$ is approximately 2~$\mu$m in all
junctions, which means that the junctions are magnetically short.
This is supported by the quasi-period of the pattern in
fig.\ref{fig:IcBHB} being close to the expected value
$\phi_0/2\lambda_Lw$ \cite{Barone}. The differential conductance
curves do not show any trace of a zero bias anomaly (ZBA), as is
expected for $0^\circ-45^\circ$ GBJs. ZBAs has been observed by
other groups in GBJs with other orientations \cite{Alff}.

The critical current is plotted as a function of applied magnetic
field for two SQUIDs in fig.{\ref{fig:IcBfig}. The result is in
qualitative agreement with theory if we assume that the SQUID
junctions have different ratios of the 1$^{\text{st}}$ and
2$^{\text{nd}}$ harmonics of the critical current. This assumption
is supported by the fairly small modulation depth (it is easy to
see from equation (\ref{eq:Iph}) that $I_c$ would go exactly to
zero in a SQUID with junctions of identical $I_{c2}/I_{c1}$).

We can fit the data to equation (\ref{eq:Iph}), if we compensate
for the residual background magnetic field and assume that we have
a small excess current (of the order of a few $\mu$A) in the
junctions. The fitting parameters again confirm that there is a
large asymmetry between the arms of the SQUIDs. Note, that the
model does not consider the flux penetration into the junctions,

\begin{figure}
\includegraphics[width=7.5cm]{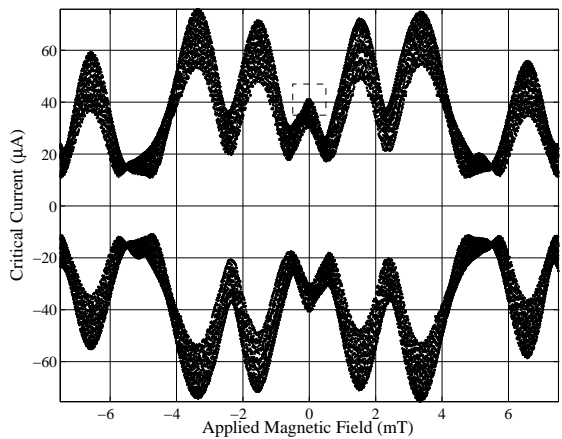}
\caption{Critical current as a function of magnetic field at 4.2K.
The dashed box indicates the area plotted in fig.
\ref{fig:IcBfig}a.} \label{fig:IcBHB}
\end{figure}

The result for fields of the order of mT is presented in
fig.\ref{fig:IcBHB} which shows the $I_c$-modulation of the SQUID
enveloped by an anomalous Fraunhofer-pattern quite similar to what
has been reported by other groups \cite{mannhart1,neils} for
$0^\circ-45^\circ$ GBJs. Note the inversion symmetry of the
pattern with respect to the origin. That the global maximum is not
in the center can be explained in several ways; it has been shown
for example that this could be due to the presence of so-called
$\pi$-loops in the junction interface \cite{smilde2002}.

\begin{figure}
\includegraphics[width=7.5cm]{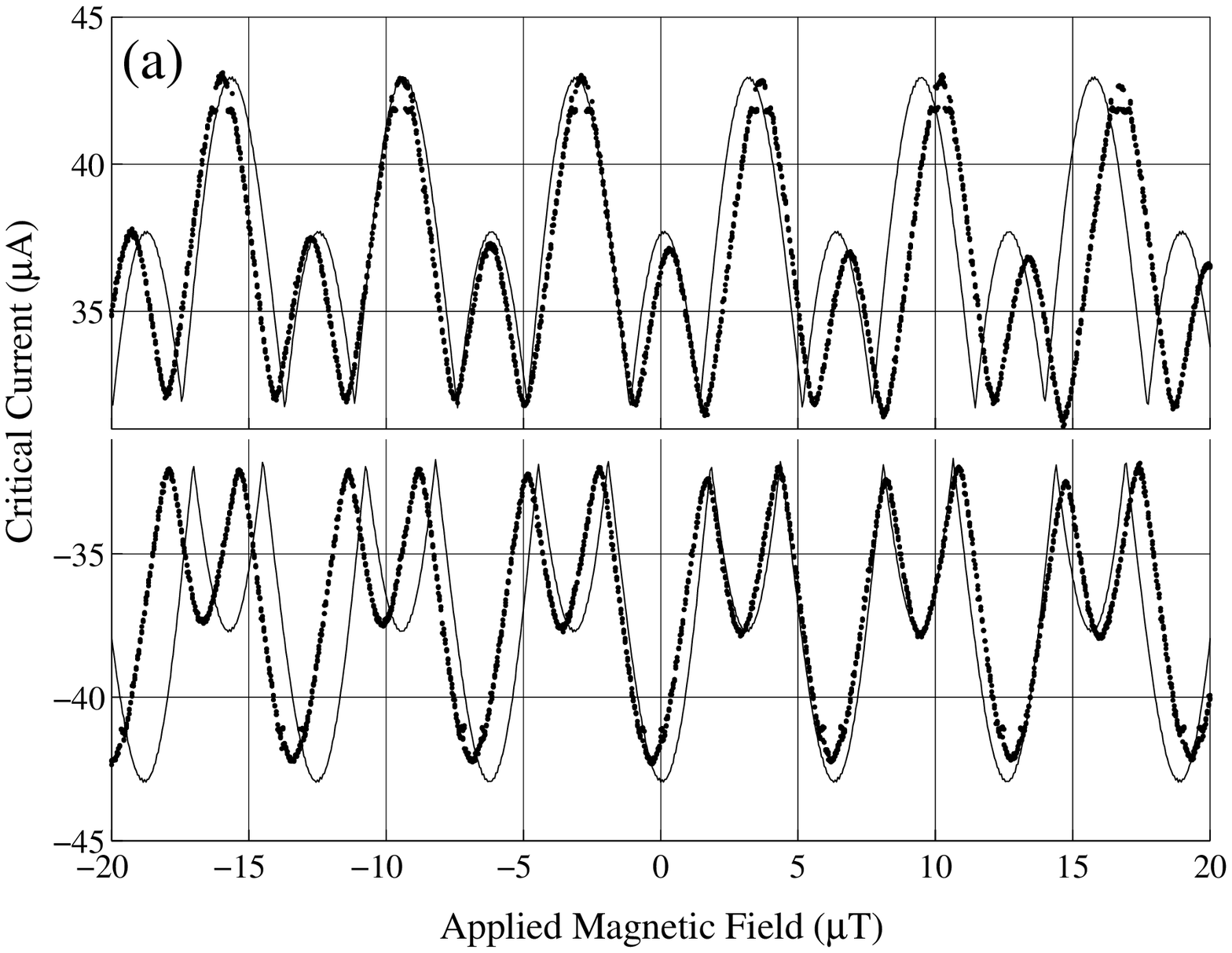}
\includegraphics[width=7.5cm]{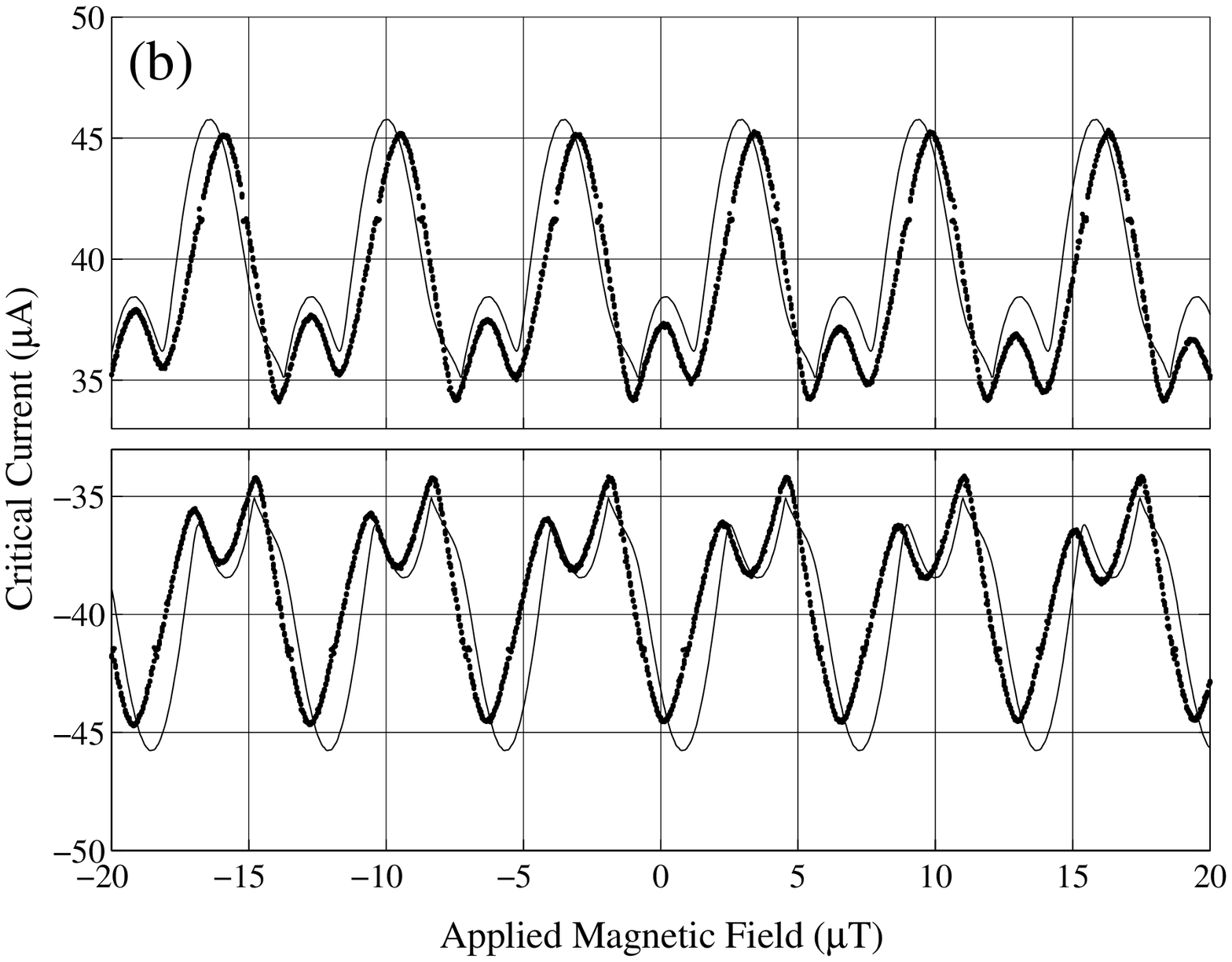}
\caption{Critical current as a function of applied magnetic field
for two different SQUIDs that are nominally identical. The solid
line represents the fitted expression. The fitting parameters are
as follows:(a) $I_{c1}^I=9$ $\mu$A, $I_{c2}^I=0.3$
 $\mu$A, $I_{c1}^{II}=3.7$ $\mu$A and $I_{c2}^{II}=22.7$ $\mu$A
 (b)$I_{c1}^I=7.8$ $\mu$A, $I_{c2}^I=3.0$
 $\mu$A, $I_{c1}^{II}=5.3$ $\mu$A and $I_{c2}^{II}=4.3$ $\mu$A . In both cases the fit has been adjusted with respect to the
 residual background field and the excess current of the
 junctions.}
\label{fig:IcBfig}
\end{figure}

Figure \ref{fig:vphi} shows the $V-B$-dependence of one of the
SQUIDs. The pattern is again field inversion-symmetric. The
overall structure is the same as in the model dependence of
fig.\ref{fig:simulations}, but there is also an additional shift
due to self-field effects, which depends on the magnitude of the
bias current and corresponds (at maximum) to a flux $\sim
0.1\Phi_0$. In a beautiful experiment a similar dependence was
recently observed by Baselmans et al in a Nb-Ag-Nb SNS junction
where current-injectors were used to change the occupation of
current-carrying states in the normal region \cite{Baselmans}.
\begin{figure}
\includegraphics[width=7.5cm]{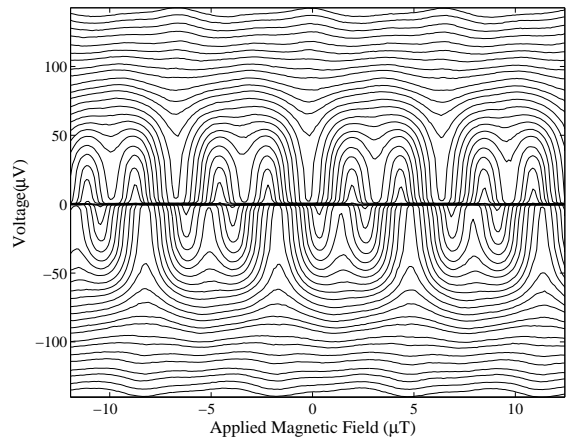}
\caption{\label{fig:vphi} Voltage modulation as a function of
applied magnetic field for the SQUID whose $I_c-B$ is shown in
fig. \ref{fig:IcBfig}a. The pattern is again inversion symmetric.
Note the sign change at 100 $\mu$V which we believe is due to a
LC-resonance in the SQUID loop.}
\end{figure}
A deviation from the model occurs at $\bar{V}$=100 $\mu$V where
the minima and maxima switch. This is probably due to an
LC-resonance in the SQUID. Taking  $L=25$ pH,  this would require
$C=0.8$ pF, which agrees with our measurements on single junctions

Remarkably, the observed offset of the $V-B$-characteristics with
respect to the {\em direction} of the bias current appears to be a
much more robust manifestation of the presence of a second
harmonic of the Josephson current, than the shape of the $I_c-B$
curves itself. We observed the shift even in SQUIDs with the
smallest junctions down to 0.5 $\mu$m wide, where the deviations
from the usual sinusoidal CPR were not obvious from the $I_c-B$
dependence.


Generally the nature of the transport through a GBJ will depend on
its transmissivity $D$. Il'ichev et al. \cite{ilichev2001} have
reported values of $D$ as high as 0.3 in symmetric
($22.5^\circ-22.5^\circ$) $dd$ junctions  as opposed to the usual
estimate for a GBJ, $D \sim 10^{-5}-10^{-2}$. Since usually
$I_c^{II}/I_c^{I} \propto D$, a high-transmissivity GBJ is
required in order to observe effects of the second harmonic.  An
estimate of the {\it average} transmissivity of our junctions
would be $\rho_{ab} l/R_NA\sim10^{-2}$ \cite{BTK} assuming $l$,
the mean free path, to be equal to 10 nm and a resistivity in the
a-b plane $\rho_{ab}$ equal to $10^{-4}$ $\Omega$cm. This is still
too low to explain the strong 2$^{\text{nd}}$ harmonic we observe.
However, it is known from, e.g., TEM-studies \cite{hilgenkamp},
that the grain-boundary is far from uniform; the properties can
significantly vary depending on the local properties of the
interface, effects such as oxygen diffusion out of the GB etc.,
which are difficult to control. It is therefore reasonable to
assume that there are many parallel transport channels through the
GB\cite{Naveh,Sarnelli}. Channels with high transmissivity
dominate the transport and might have $D\sim0.1$ even though the
{\em average} transmissivity is much lower. This is also
consistent with the fact that most of our SQUIDs seem to be highly
asymmetrical which is to be expected if the distribution of
channels is random. The ratios of $I_c^I$ and $I_c^{II}$ can vary
as much as ten times between two junctions in the same SQUID, even
though the fluctuations of the {\it total} $I_c$ from sample to
sample are much smaller. It is also clear from general
considerations that a high value of $I_c^{II}$ {\it excludes} a
high value of $I_c^{I}$, since the 2$^{\text{nd}}$ harmonic
usually dominates if the odd harmonics of the supercurrent are
cancelled by symmetry \cite{barash}.

Recent studies of $0^\circ-45^\circ$ GBJs have demonstrated that
the SQUID-dynamics can be altered by the d-wave order parameter in
YBCO \cite{chesca}. It is however important to point out that our
results {\it do not} directly relate to e.g. tetracrystal
$\pi$-SQUID experiments; the latter crucially depend on having one
$\pi$-junction with negative critical current, but still only the
first harmonic present in $I_c(\phi)$. Our SQUIDs have a
conventional geometry, but unconventional  current-phase
relations.

One explanation for the pronounced effects of the 2$^{\text{nd}}$
harmonic could be that relatively large sections of the interface
are highly transparent and have a low degree of disorder. This in
turn could be related to our fabrication scheme which seems to
preserve the integrity of the barrier. This makes feasible their
applicability in the quantum regime and supports our expectations
that quantum coherence can be observed in this kind of structures.

To summarize, we have observed very pronounced 2$^{\text{nd}}$
harmonic in the current-phase relation of a 'conventional' YBCO
dc-SQUID with $0^\circ-45^\circ$ grain boundary junctions. It has
strongly influenced the SQUID dynamics. All details of the SQUID
behavior were explained within a simple model of a dd-junction
with relatively high transparency. We believe that these effects
are important for better understanding of HTS Josephson junction
and SQUIDs.


\begin{acknowledgments}
Discussions with Evgenii Il'ichev, Alexander Golubov, Tord
Claeson, and John Gallop are gratefully acknowledged. The work is
in part supported by The Board for Strategic Research (SSF) via
the "OXIDE" program, the Science Research Council, and the "Fonds
québécois de la recherche sur la nature et les technologies". The
processing work is done at the MC2 process lab at Chalmers
University of Technology.
\end{acknowledgments}

\bibliography{anomalous_CPR}

\end{document}